\documentclass{ws-ijmpc}

\begin{document}

\markboth{C. Zou et al.}
 {Lattice--BGK Model Collision Consistency}

\catchline{}{}{}{}{}

\title{
 THE COLLISION CONSISTENCY OF LATTICE-BGK MODEL FOR SIMULATING
 RAREFIED GAS FLOWS IN MICROCHANNELS
 \footnote{This work is supported by the National Key Basic Research and
 Development Program of China (Grant No. 2004CB217703 and 2006CB705800).}
}

\author{CHUN ZOU, ZHI-WEI TIAN, HONG-JUAN LIU and CHU-GUANG ZHENG}

\address{State Key Laboratory of Coal Combustion,\\
 Huazhong University of Science and Technology,\\
 Wuhan Hubei 430074, P.R. China\\
 zouchun@mail.hust.edu.cn (C Zou)\\
 zwtiantian@gmail.com (ZW Tian)}

\maketitle

\begin{history}
 \received{\today}
 \revised{Day Month Year}
\end{history}

\begin{abstract}
The collision consistency between the BGK collision model equation and
lattice-BGK (LBGK) model is proposed by researching the physical significance
of the relaxation factor $\tau$ in LBGK model. For microscalar flow in which
the continuum hypothesis is not still satisfied, the collision consistency
$\tau=1.0$ should be ensured when using the LBGK model for simulating
microflows. The results of simulating microchannel Poiseuille flow with
constant pressure gradient under collision consistency by using LBGK model are
well consistent with the analytical solutions, and the accuracy of these
results is three or four orders of magnitude higher than those that don't
satisfy the collision consistency.

\keywords{Collision consistency; LBGK model; Rarefied gas flow;
 Numerical simulation.}
\end{abstract}

\ccode{PACS Nos.: 04.60.Nc, 47.45.-n, 47.61.-k}

\section{Introduction}

Micro-electro-mechanical systems (MEMS) technology has been developed rapidly
in recent years.\cite{mems98}\~{}\cite{mf}  The characteristic length scale of
MEMS is typically the order of microns, and the ratio of the mean free path to
the characteristic dimension (i.e. Knudsen number $Kn$) can not be negligible.
The dynamics associated with microchannels can thereby exhibit rarefied
phenomena and compressibility effects. The former is the emergence of a slip
velocity at the wall boundary in microchannel flows. The latter is the
significant nonlinear pressure drop of gas flowing in a long microchannel.
Because conducting experiments in micrometer-size is a big challenge, numerical
simulations of MEMS become very important tools of investigation. But the
methods commonly used in simulation of MEMS, such as molecular dynamics (MD),
direct simulation Monte--Carlo (DSMC)\cite{dsmc98} and direct numerical
simulation of Boltzmann equation, usually requires a tremendous amount of
computer time and memory. Recently, for its intrinsic kinetic nature, the
Lattice Boltzmann method (LBE) has been become an attractive method for
simulation of microscalar flows where both microscopic and macroscopic
behaviors are coupled.\cite{el89,el92}

Form its birth nearly 20 years ago (1988)\cite{prl88}, the lattice Boltzmann
method (LBM)\cite{ben92}\~{}\cite{liu} has met with significant success for the
numerical simulation of a large variety of fluid flows, and has emerged as an
alternative numerical technique for simulating fluid flows. The LBE method
usually solves the Bhatnagar-Gross-Krook (BGK) model equation, a simplified
model Boltzmann equation, on a discrete lattice, which is usually named the
lattice-BGK (LBGK) model.\cite{lb98,liu} Since theoretical connections between
the LBGK model and the Boltzmann equation\cite{he97,shan98} and the preliminary
link between LBGK model and the Burnett-type equations\cite{qian} were
established, the LBE model can be valid for rarefied gas flow provided the Mach
number is small. Moreover, since the standard BGK equation is able to simulate
highly non-equilibrium gas flows, the LBGK model should also be applicable to
rarefied microflows theoretically.\cite{nie}\~{}\cite{tian5}

One of the most important issues of LBGK model for simulating microchannel
flows is the introduction of $Kn$ number into LB models. The popular treatment
is the construction of the relationship between $Kn$ and relaxation time $\tau$
of LB models. Nie \emph{et al.} used the explicit LBE formulation and related
the nondimensional relaxation time $\tau_e$ to Knudsen number $Kn$ as: $Kn=
\alpha (\tau_e-0.5)/ \rho H$ for a microchannel of height $H$ and gas density
$\rho$. The factor 0.5 attributes to the explicit treatment of the collision
term and $\alpha$ is chosen to best match the simulated mass flow rate with
experiments.\cite{nie} Later, Lim \emph{et al.}\cite{lim} proposed a different
relation between $Kn$ and $\tau_e$ for the explicit LBE formulation without the
correction factor of 0.5 as in Nie \emph{et al.}. Their $Kn$ for a long
microchannel was defined as $Kn=\left( \delta x \tau_e /H \right) /\left( p_0/p
\right)$, where $p$ and $p_0$ are the local pressure and pressure at the outlet
of the microchannel, respectively. Zhang \emph{et al.} gave different
definitions of $\tau_e$ with different constant factors among various lattice
models.\cite{pre05} Recently, Lee \emph{et al.} proposed a definition of $\tau$
about fully implicit LBE.\cite{pre05lee}

Another one is the implementation of boundary condition. Nie \emph{et al.} used
the half bounce-back rule for the slip effect at the surface. Lim \emph{et al.}
used a specular reflection model to generate slip effect. Zhang \emph{et al.}
adopted the Maxwellian scattering kernel to address the gas molecule and
surface interactions with an accommodation coefficient $\alpha$. Lee \emph{et
al.} proposed a wall equilibrium condition according to the assumption of rough
surface on the characteristic length of gas molecules.

For the former issue, the previous research is still related to the framework
of Navier--Stokes equations. This is because the expression of the kinetic
viscosity coefficient derived from the LB equation recovering the
Navier--Stokes equations, is still adopted in the makeup of the relationship of
$Kn$ and $\tau$. Furthermore, when $Kn$ approaches zero, the Boltzmann equation
can be reduced to the Navier--Stokes equation. But all the existing LBE for
microscale flow can not be reduced to the Navier--Stokes equation because of
using the relationship between $Kn$ and $\tau$.

The previous researchers always focus on recovering the Navier--Stokes
equations precisely from the LBGK model and ignoring the deviation of the LBGK
model and Boltzmann equation. This deviation may not do any effect in
simulation of macroscale flows. However, the character scale of microflows is
much smaller. In such a case, keeping the collision consistency of the LBGK
model and Boltzmann equation should be paid special attention. This collision
consistency requires the collision frequency of LBGK model is equal to that of
BGK model equation. With the further study on the relaxation factor $\tau$, we
found that these two collision frequency are equivalent when $\tau=1.0$ and the
simulation results are the most approximate. This principle is called collision
consistency of LBGK model.

On the other hand, the boundary treatment is based on the Maxwell slip model
straightforwardly. There is not any adjustable parameter in our model and is
similar to the Newmann boundary condition in some degree. The implement is not
related to the Navier--Stokes equations any more, so it can be applied in
simulating microflows not only in slip regime but also in transition regime.

\section{Lattice-BGK Equation and Collision Consistency}

To begin with, the BGK equation is:
\begin{equation}\label{bgk}
 \frac{\partial h}{\partial t} +\textbf{c}\cdot \nabla h
 =-\frac{1}{\tau_0} \left( h-h^{eq} \right)
\end{equation}
here $h=h(x,c,t)$ and $h^{eq}=h^{eq}(x,c,t)$ are the molecular distribution
function and equilibrium distribution function, respectively.
\begin{equation}\label{heq}
 h^{eq}= \frac{\rho}{m} \left( \frac{m}{2\pi k_B T} \right)^{D/2}
 \exp \left[ -\frac{m(\textbf{c}-\textbf{u})^2}{2k_B T} \right]
\end{equation}
where $\textbf{c}$ and $\textbf{u}$ are molecular velocity and macroscopical
velocity respectively; $\tau_0$ is the collision time; $1/\tau_0$ represents
the collision frequency: $1/\tau_0 = \bar{c}/\lambda$. where $\bar{c}$ is
molecular mean velocity; $\lambda$ is molecular mean free path.

LBGK model equation is:
\begin{equation}\label{lbgk}
 \frac{\partial f_i}{\partial t} +\textbf{c}_i\cdot \nabla f_i
 =-\frac{1}{\tau_0} \left( f_i-f_i^{eq} \right)
\end{equation}
here $f_i$ is particle distribution function:
\begin{equation}\label{fi}
 f_i= \omega_i \left[1+ \frac{(\textbf{c}_i \cdot \textbf{u})}{c_s^2} +
 \frac{(\textbf{c}_i \cdot \textbf{u})^2}{2c_s^4} - \frac{\textbf{u}^2}{2c_s^2}\right]
\end{equation}
$\textbf{c}_i$ is lattice discrete velocity and $\omega_i$ is weighing factor,
$\tau_0$ is the collision time.

LBGK equation (\ref{lbgk}) can be further discretized in space and time. The
completely discretized form of Eq.~(\ref{lbgk}), with the time step $\Delta t$
and space step $\Delta x =c_i \Delta t$, is:
\begin{eqnarray}\label{fievol}
 {f_i}(\textbf{x}+ \textbf{c}_i \Delta{t}, t+\Delta{t}) &-& {f_i}(\textbf{x}, t)
 = -\frac{1}{\tau} \left[{f_i}(\textbf{x},t) - f_i^{eq} (\textbf{x}, t)\right]
\end{eqnarray}
where $\textbf{x}$ is a point in the discretized physical space, and it is
worth notice that $\tau$ commonly considered as a nondimensional relaxation
time, $\tau={\tau_0}/{\Delta t}$, represents the collision consistency of LBGK
model by the analysis of the physical signification of $\tau_0$ and $\Delta t$.

The above discrete LBGK equation is usually solved in two steps: collision step
and streaming step. So the physical process in LBGK model can be regarded as a
large number of particles take place one collision, through the distance
$\Delta x$ and the time $\Delta t$. Therefore, $\Delta x$ and $\Delta t$ can be
regarded as the mean free path and the collision time in the LBGK model,
respectively, and $1/\Delta t$ represents the collision frequency in the LBGK
model. Consequently, $\tau={\tau_0}/{\Delta t}$ means the ratio of the
collision frequency of the LBGK model to that of the BGK model. From the gas
dynamics, the collision frequency $\tau_0$ is in direct proportion to the
density, and in dependence on temperature, but independence of molecular
velocity. In order to furthest approach to BGK model, the collision frequency
of the LBGK model should be equal to that of the BGK model, i.e.
$\tau={\tau_0}/{\Delta t}=1$. Especially in the microflows ($Kn>10^{-3}$), the
collision consistency must be much more ensured for the collision frequency of
molecule in the microflows is far lower than that in the macroscopical flows
($Kn<10^{-3}$).

It must be emphasized that LBGK model is between the BGK equation and
Navier--Stokes equations. We formerly attached importance to recover the LBGK
to Navier--Stokes equations in the simulation with LBGK model, but now we
should pay sufficient attention to the consistency of the LBGK model with the
BGK equation, especially in simulation of the microflows.

\section{Knudsen Number, Lattice Number and Boundary Condition}

In the LBGK model, $\Delta t$ is defined as:
\begin{equation}\label{dt}
 \Delta t = \frac{\Delta x}{\textbf{c}}
\end{equation}
and in the BGK model, $\tau_0$ is defined as:
\begin{equation}\label{tau0}
 \tau_0=\frac{\lambda}{\bar{c}}
\end{equation}

Using Eq.~(\ref{dt}) and Eq.~(\ref{tau0}), the relaxation factor $\tau$ can be
determined as:
\begin{equation}\label{tau}
 \tau=\frac{\Delta t}{\tau_0} =\frac{\bar{c}}{c} \frac{\Delta x}{\lambda}
 =\frac{c}{\bar{c}} \frac{1}{N_H Kn}
\end{equation}
where $N_H$ is the number of the lattice in the characteristic length $H$,
$H=\Delta x N_H$ .

Regarding $\bar{c}$ approximately equal to $c$\cite{pre05lee}, Eq.~(\ref{tau})
can be simplified:
\begin{equation}\label{taus}
 \tau=\frac{1}{N_H Kn}
\end{equation}

According the collision consistency, the relation of $Kn$ to the number of the
lattice is:
\begin{equation}\label{n}
 N_H=\frac{1}{Kn}
\end{equation}
From Eq.~(\ref{n}), the number of lattice should be in inverse proportion to
the Kn, rather than be determined randomly.

Under the collision consistency, Eq.~(\ref{fievol}) can be simplified as:
\begin{equation}\label{fis}
 {f_i}(\textbf{x}+ \textbf{c}_i \Delta{t}, t+\Delta{t}) = f_i^{eq} (\textbf{x}, t)
\end{equation}
The distribution function $f_i$ evolved in the lattice actually is the
equilibrium distribution function $f_i^{eq}$. Hence, if the velocity on the
boundary can be determined, the distribution function of the boundary can also
be determined.

To the slip flow regime, the macroscopical slip-velocity on the boundary can be
obtained by:
\begin{equation}\label{us}
 u_s=  \frac{2-\sigma}{\sigma} \lambda \left(\frac{\partial u}{\partial y} \right)_{wall}
\end{equation}
where $u_s$ is the boundary slip-velocity; $\sigma$ is accommodation factor.
Using second order difference formula, Eq.~(\ref{us}). Can be discretized as:
\begin{equation}\label{uss}
 u_s=  \frac{Kn \left( 4u_1 - u_2 \right)}{2\Delta x +3Kn}
\end{equation}
where $u_1$ and $u_2$ are the velocity of the first and the second point
adjacent the wall, respectively.

\section{Numerical Results and Discussion}

In this paper, we study microscalar gas Poiseuille flows in a constant external
pressure gradient in order to validate the collision consistency. The
compressibility effect becomes negligible and only the rarefaction effect is
accounted for. We study three cases with different $Kn$ number value
($Kn$=0.02, 0.05, 0.1), and the number of grids adopted in the simulations is
chosen according to the collision consistency. Results will be compared with
the analytical results and the results of more fine grids and relaxation factor
chosen according to the Eq.~(\ref{taus}).

\subsection{Slip Flows}

\begin{figure}[tb]
\centering
\includegraphics[width=0.75\textwidth]{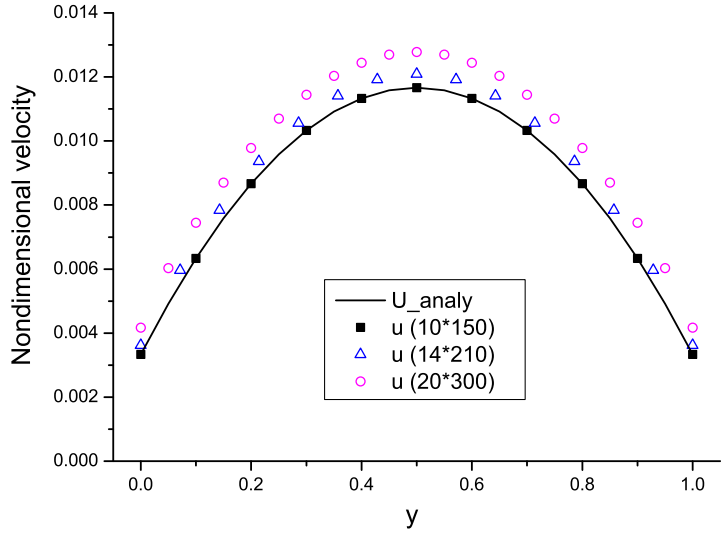}
\caption{The comparison of velocity distribution of micro Poiseuille flow
($Kn=0.1$).}\label{01}
\end{figure}

In case 1, when $Kn$=0.1, and the pressure gradient $dp/dx=6.6667 \times
10^{-2}$. According to the collision consistency, relaxation factor $\tau=1.0$,
and the grid numbers are $10\times 150$. From the calculating result (see
Fig.~\ref{01}), we can see the results agree closely with the analytical
results when $\tau=1.0$, and the relative error is between $1.43-1.54 \times
10^{-8}$. When $\tau=1/1.4$ and $\tau=0.5$, the fine grids are increased, but
the simulative relative error is very big because of departing away from the
collision consistency.

\begin{figure}[tb]
\centering
\includegraphics[width=0.75\textwidth]{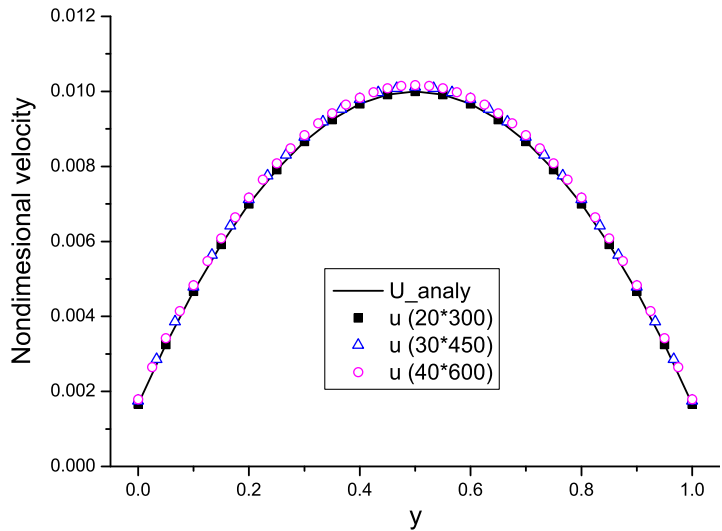}
\caption{The comparison of velocity distribution of micro Poiseuille flow
($Kn=0.05$).}\label{005}
\end{figure}

In Figure \ref{005}, when $Kn=0.05$ and the pressure gradient $dp/dx=6.6667
\times 10^{-2}$, relaxation factor $\tau=1.0$ according to the collision
consistency, we can see the simulating results agree closely with the
analytical results, and the relative error is between $1.53-1.61 \times
10^{-6}$. We increased the grids by 1.5 and 2 times and got the relaxation
factor according to Eq.~(\ref{taus}), but the relative error was still very
large.

\begin{figure}[tb]
\centering
\includegraphics[width=0.75\textwidth]{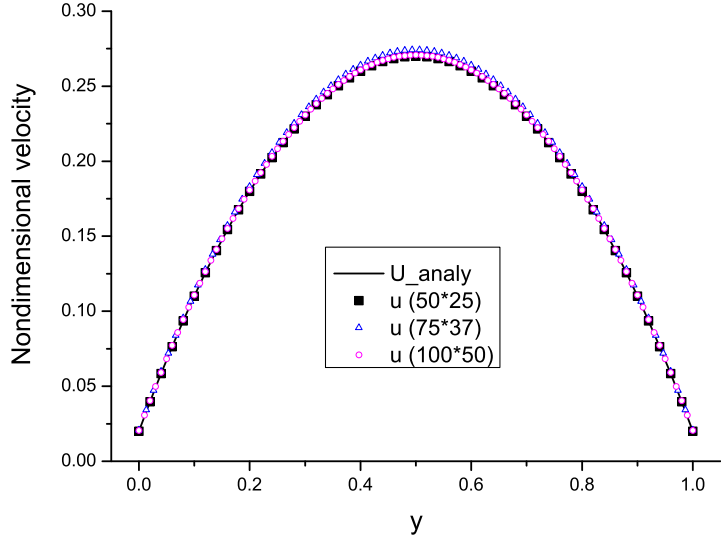}
\caption{The comparison of velocity distribution of micro Poiseuille flow
($Kn=0.02$).}\label{002}
\end{figure}

Figure \ref{002} shows the comparison of calculating results obtained in
different grids and relaxation factors when $Kn=0.02$ and the pressure gradient
is 2.0. As can be seen from the graph, the results accord closely with the
analytical results, but in the other two cases, the accuracy of the simulating
results decreased instead of increasing although the grids increased 1.5 and 2
times, about four times less than that of when $\tau=1.0$.

\begin{table}[tb]
 \tbl{The relaxation factors and relative calculation error in different situations.}
{
\begin{tabular}
 {@{}ccccccc@{}}
 \toprule
 $Kn$ & Grid & $dp/dx$ & Relaxation & Relative error & Relative error \\
 & & & factor & (Min)& (Max) \\
 \colrule
 0.1\hphantom{0} & $10\times 150$ & $6.6667\times 10^{-2}$ & 1.0
   & $1.43\times 10^{-8}$ & $1.54\times 10^{-8}$ \\
 0.1\hphantom{0} & $14\times 210$ & $6.6667\times 10^{-2}$ & 1/1.4
   & 0.0362 & 0.0857 \\
 0.1\hphantom{0} & $20\times 300$ & $6.6667\times 10^{-2}$ & 0.5
   & 0.096\hphantom{0} & 0.102\hphantom{0} \\
 0.05 & $20\times 300$ & $6.6667\times 10^{-2}$ & 1.0
   & $1.53\times 10^{-6}$ & $1.61\times 10^{-6}$ \\
 0.05 & $30\times 450$ & $6.6667\times 10^{-2}$ & 1/1.5
   & 0.012\hphantom{0} & 0.05\hphantom{00} \\
 0.05 & $40\times 600$ & $6.6667\times 10^{-2}$ & 0.5
   & 0.0167 & 0.075\hphantom{0} \\
 0.02 & $50\times 25$ & 2.0 & 1.0 & $2.38\times 10^{-6}$ & $2.92\times 10^{-6}$ \\
 0.02 & $75\times 37$ & 2.027 & 1/1.5 & 0.0021 & 0.02\hphantom{00} \\
 0.02 & $40\times 600$ & 2.0 & 0.5 & 0.003\hphantom{0} & 0.03\hphantom{00} \\
 \botrule
\end{tabular} \label{tab}
}
\end{table}

All the calculating errors of different grids and relaxation factors in these
three cases are shown in Table~1. For micro Poiseuille flow, the best
calculating results are obtained since we adopted the relaxation factors in
according to the collision consistency. So the simulating precision is 3 to 4
times more than that of the other results. From Table~1, we can see, the errors
are almost the same under the same relaxation factors. This also illustrates
the influence of the boundary error on the whole flow simulating error.

\subsection{Higher $Kn$ Flows}

We now investigate the simulation for higher $Kn$ number in this subsection.
The ratio of the microchannel length $L$ to height $H$ is chosen as 100, and
a 2000$\times$20 regular grid is applied, which has been verified by
grid-dependence. We focus on the streamwise velocity profile $U$ at the exit
normalized by the centerline maximum velocity and the pressure deviation
$P'=(p-p_l)/p_o$ from the linear distribution $p_l= p_o+ (p_i-p_o)(1-x/L)$.
Note that we choose the pressure deviation $P'$ rather than the pressure
distribution itself in order to show the difference more clearly.

\begin{figure}[tb]
\centering
\includegraphics[width=0.75\textwidth]{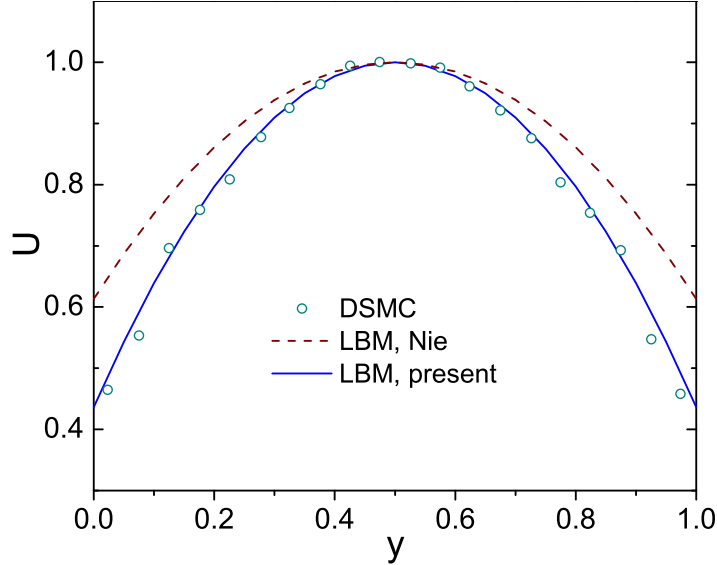}
\caption{Normalized velocity profile at microchannel outlet
($Kn$=0.194).}\label{U194}
\end{figure}

\begin{figure}[tb]
\centering
\includegraphics[width=0.75\textwidth]{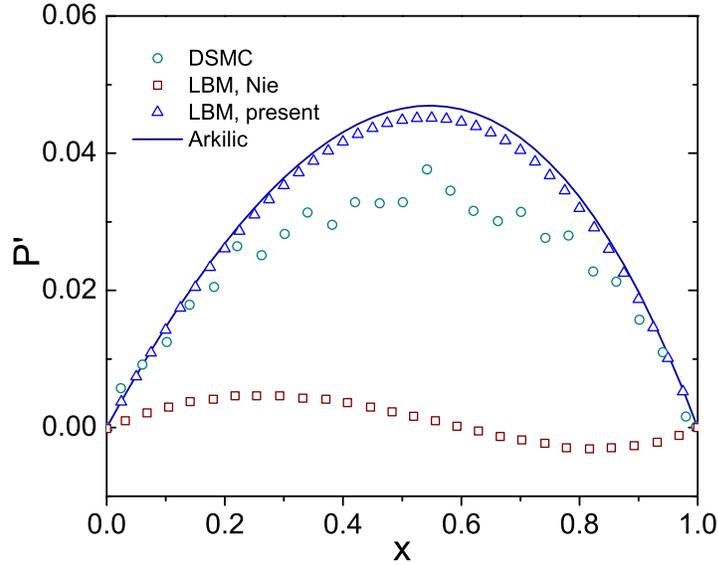}
\caption{Pressure deviation $P'$ along the microchannel
($Kn$=0.194).}\label{dP194}
\end{figure}

Fig.~\ref{U194} shows the velocity profile at the outlet of the microchannel
with $Kn$=0.194. Compared with the reliable DSMC result, Nie \emph{et~al.}
over-predicted the slip velocity obviously and the velocity distribution is
more higher in nearly the whole outlet cross-section. In our simulation, both
the slip velocity and exit velocity distribution are in consistent with the
results of DSMC method. Moreover, pressure deviation $P'$ is shown in
Fig.~\ref{dP194}. The \emph{qualitative} difference with DSMC method has been
founded in Nie \emph{et~al.}'s numerical result. Our present result is very
closely to the analytical solution by Arkilic \emph{et~al.}. And the deviation
from the DSMC result appears only in the middle part along the microchannels.

\section{Conclusion}

From our study, the physical meaning of the relaxation factor $\tau$ in LBGK
model is: the ratio of the collision frequency in lattice system and in the BGK
model. LBGK model situates between BGK model and macro Navier-Stokes equations
so the collision consistency needs analyzing when LBGK model is dispersed from
the BGK collision equation. This is very essential to adopting LBGK model to
simulate the microflow because the continuum assumption may be break down in
the microflow. Therefore we should assure the same collision frequency in LBGK
model and BGK model, that is collision consistency.

Under collision consistency, it is simple to deal with the boundary and
determine the relation factor. Adopting the collision consistency principle
$\tau=1.0$, the results of the simulating Poiseuille flow accord closely with
the analytical results. Without it, the precision of simulating results is
three to four times less even if the greater grids are adopted.



\begin{thebibliography}{00}

 \bibitem{mems98} C. M. Ho and Y. C. Tai,
 {\it Annu. Rev. Fluid Mech.} {\bf 30}, 579 (1998).

 \bibitem{be02} D. J. Beebe, G. A. Mensing and G. M. Walker,
 {\it Annu. Rev. Biomed. Eng.} {\bf 4}, 261 (2002).

 \bibitem{mf} G. E. Karniadakis and A. Beskok,
 {\it Micro Flows: Fundamentals and Simulation} (Springer, New York, 2001).

 \bibitem{dsmc98} E. S. Oran, C. K. Oh and B. Z. Cybyk,
 {\it Annu. Rev. Fluid Mech.} {\bf 30}, 403 (1998).


 \bibitem{el89} F. J. Higuera, S. Succi and R. Benzi,
 {\it Europhys. Lett.} {\bf 9}, 345 (1989).

 \bibitem{el92} Y. H. Qian and S. A. Orszag,
 {\it Europhys. Lett.} {\bf 21}, 255 (1993).

 \bibitem{prl88} G. McNamara and G. Zanetti, {\it Phys. Rev. Lett.} {\bf 61}, 2332 (1988).


 \bibitem{ben92} R. Benzi, S. Succi and M. Vergassola,
 {\it Phys. Report.} {\bf 222}, 145 (1992).

 \bibitem{lb98} S. Chen and G. D. Doolen,
 {\it Annu. Rev. Fluid Mech.} {\bf 30}, 329 (1998).

 \bibitem{liu} H. J. Liu, C. Zou, B. C. Shi, Z. W. Tian, L. Q. Zhang, and C. G. Zheng,
 {\it Int. J. Heat Mass Tran.} {\bf 49}, 4672 (2006).


 \bibitem{he97} X. He and L. S. Luo, {\it Phys. Rev. E} {\bf 55}, R6333 (1997).

 \bibitem{shan98} X. Shan and X. He, {\it Phys. Rev. Lett.} {\bf 80}, 65 (1998).

 \bibitem{qian} Y. H. Qian and Y. Zhou, {\it Phys. Rev. E} {\bf 61}, 2103 (2000).


 \bibitem{nie} X. B. Nie and G. D. Doolen, {\it J. Stat. Phys.} {\bf 107}, 279 (2002).

 \bibitem{lim} C. Y. Lim, C. Shu, X. D. Niu and Y. T. Chew,
 {\it Phys. Fluids} {\bf 14}, 2299 (2002).

 \bibitem{tang} G. H. Tang, W. Q. Tao, and Y. L. He,
 {\it Int. J. Modern Phys. C} {\bf 15}, 335 (2004).

 \bibitem{shu} X. D. Niu, C. Shu, and Y. T. Chew,
 {\it Int. J. Modern Phys. C} {\bf 16}, 1927 (2005).

 \bibitem{pre05} Y. Zhang, R. Qin and D. R. Emerson,
 {\it Phys. Rev. E} {\bf 71}, 047702 (2005).

 \bibitem{pre05lee} T. Lee and C. L. Lin,
 {\it Phys. Rev. E} {\bf 71}, 046706 (2005).

 \bibitem{tian5} Z. W. Tian, C. Zou, Z. H. Liu, Z. L. Guo, H. J. Liu, and C. G. Zheng,
 {\it Int. J. Modern Phys. C} {\bf 17}, 603 (2006).

\end{thebibliography}
\end{document}